# A STUDY ON BINARY ASTEROID SYSTEM DEFLECTION

**Lucas G. Meireles**[(1)], **Antônio F. B. de A. Prado**[(1)], **Maria Cecília Pereira**[(2)], **Cristiano F. de Melo**[(2)]

[(1)] National Institute for Space Research, Av. dos Astronautas, 1758, Jd. Granja, São José dos Campos, SP, Brazil, 12227-010, meireleslg@gmail.com, antonio.prado@inpe.br

[(2)] Federal University of Minas Gerais, Av. Pres. Antônio Carlos, 6627, Pampulha, Belo Horizonte, MG, Brazil, 31270-901, cecilia@demec.ufmg.br, cristiano.fiorilo@demec.ufmg.br

The study of asteroids, its composition and trajectories, has been a persistent interest in the space exploration community. In addition, they are also perceived as a great threat to life on Earth, considering the possibility of an impact with our planet. A considerable portion, around 15%, of the asteroid population are believed to be part of a double or triple asteroid system. From a planetary defense mission perspective, the deflection of the smaller body of a binary system is interesting given the smaller velocities of its orbit around the primary body, facilitating any deflection of its trajectory measurements. With that in mind, this study proposes an analysis of small increments in the velocity of secondary bodies of different binary asteroid systems and its effect in the trajectory of the system. A system with similar properties to the Didymos binary asteroid was considered, taking into account the relevance of the recently launched Double Asteroid Redirection Test (DART) mission, by NASA. Some characteristics of the system were modified such as the mass ratio between the bodies, their initial distance, the magnitude and direction of the velocity increment. Numerical integration of a 3-body problem were performed for the duration of an orbital revolution and the trajectories were compared to reference trajectories where there were no velocity increment applied to the secondary body. Results take into account possible inelastic collision between the bodies and even escape scenarios from the secondary body. Altogether, it is possible to verify noticeable deviation in some of the trajectories simulated for the main body of the system.

## 1. Introduction

The study of asteroids, its composition and trajectories, has been a persistent interest in the space exploration community. Within them lies secrets to the formation of our Solar System. For instance, Trojan Asteroids were said to be "the fossils of planet formation" by Harold Levison of the Southwest Research Institute in Boulder, Colorado. A number of missions were deployed by different space agencies with the goal of analyzing various characteristics of asteroids such as the Lucy Mission [1] and OSIRIS-REx [2] from the National Aeronautics and Space Administration (NASA), Rosetta [3, 4] from the European Space Agency (ESA), Hayabusa [5] and Hayabusa2 [6, 7] from the Japan Aerospace Exploration Agency (JAXA).

In addition, asteroids are also perceived as a great threat to life on Earth, considering the possibility of an impact with our planet. In worst case scenarios, they would be capable of unveiling even mass extinction events. In this manner, it is interesting not only to know their properties but also to be able to control their trajectories, even to a slightest extent. With that in mind, NASA set out to launch the Double Asteroid Redirection Test (DART) [8], to test the extent to which the trajectory of an asteroid can be deflected by a human vehicle, within other goals. Aiming for the Didymos and Dimorphos binary asteroid system, the DART mission will project its kinetic impactor onto the smaller body Dimorphos and study the amount of momentum transfer along with the deflection of the system's trajectory.

It is from that mission's objectives that this study proposes an analysis of small velocity increments in the secondary body of a binary asteroid system and its effect on the systems orbital properties.

## 2. Methodology

The simulations consist in incrementing the velocity of the secondary body of a binary asteroid system with different magnitudes and with a variety of different initial conditions. A numerical integration of a 3-body problem is then performed for the duration of an entire revolution around the Sun and the final position of the main body is compared to its final position in a simulation with zero velocity increment. Note that any collision that occurred between the two bodies was considered to be an inelastic collision.

This binary system was considered to have its main body in a circular orbit with semi-major axis of 1 A.U. and an inclination of $0°$ of its plane of orbit. The secondary body was also considered to be in a circular orbit around the main body. A total mass of $m_T = 5.4 \times 10^{11}$ kg was considered [9], taking from inspiration the Didymos and Dimorphos system, the target of DART mission. A total of three cases were invetigated.

*2.1. Case 1*

Three characteristics were modified, in order to analyze the difference in their influence over the expected results:

- Increment in velocity: $\Delta v = [0, 1, \ldots 10] * \frac{0.01}{1-m_r}$ mm/s

- System mass ratio: $m_r = [0.8, 0.81, \ldots 0.99]$

- Initial distance between bodies: $d_0 = [10, 15, \ldots 40]$ km

The increment in velocity was given in the direction of the movement. In order to maintain equivalent gains in the linear quantity of motion for simulations of systems of different mass ratio $m_r$, the magnitude of the increment was multiplied by a normalizing factor $\frac{0.01}{1-m_r}$. This was necessary because the total mass of the system $m_T$ was always the same, so a variation of $m_r$ resulted in a variation of the mass of the secondary body. In turn, the same magnitude in the increment of velocity resulted in different gains in momentum.

In addition, a 2-Body Problem base scenario with equivalent gains in momentum was performed. This served as a basis to compare the same increment in momentum of the system but implemented on the main body, without the gravitational interaction from the secondary body.

- 2-Body Problem
- Increment in velocity: $\Delta v = [0, 1, ... 10] * \frac{0.01}{m_A}$ mm/s
- Asteroid mass: $m_A = [0.8, 0.81, ... 0.99] * m_T$

*2.2. Case 2*

The same three characteristics from Case 1 were modified in a slightly different manner, with the addition of an extra variable: the direction of the increment in velocity.

- Increment in velocity (magnitude): $\Delta v = [0, 1, ... 10] * \frac{0.01}{1-m_r}$ mm/s
- System mass ratio: $m_r = [0.8, 0.81, ... 0.99]$
- Increment in velocity (direction): $\angle \Delta v = [0, 15, ... 345]°$
- Initial distance between bodies: $d_0 = [10, 40]$ km

It is important to note that an increment of velocity with an angle of $0°$ is given at the direction of movement of the secondary body. An angle of $180°$ means an increment in the direction opposite to the movement.

*2.3. Case 3*

In this case, a system with an elliptical orbit around the Sun was considered, with a semi-major axis equal to $2$ A.U. and an eccentricity of $0.5$. A total of four characteristics were modified:

- Increment in velocity: $\Delta v = [0, 1, ... 10] * \frac{0.01}{1-m_r}$ mm/s
- System mass ratio: $m_r = [0.8, 0.81, ... 0.99]$
- Initial system's true anomaly: $f_0 = [0, 15, ... 345]°$
- Initial distance between bodies: $d_0 = [10, 40]$ km

## 3. Results

The results consist of a brief analysis of the difference in the final heliocentric position of the main body ($\Delta r_{12}$) of the asteroid system, after a complete simulation. In other words, how much a velocity increment of the secondary body resulted in a deflection of the trajectory of the primary body. The results are displayed in three dimensional plots.

## 3.1. Case 1

From Figures 2 to 4, the X-axis indicates the variation of the increment in velocity ($\Delta v$), the Y axis the variation in the initial distance between the bodies ($d_0$) and the Z axis the difference in the final position of the primary asteroid ($\Delta r_{12}$) compared to a reference simulation with $\Delta v = 0$ mm/s. Different mass ratios ($m_r$) are displayed along different subplots.

Additionally, Figure 1 displays the results for the 2-Body Problem scenario. It is possible to access that the best case values of deflection are approximately 10 km, for the largest magnitudes of velocity increment. An increase in $m_r$ results in lower $r_{12}$ values, demonstrating that asteroids with greatest masses will not suffer larger deflections given the same gain in momentum.

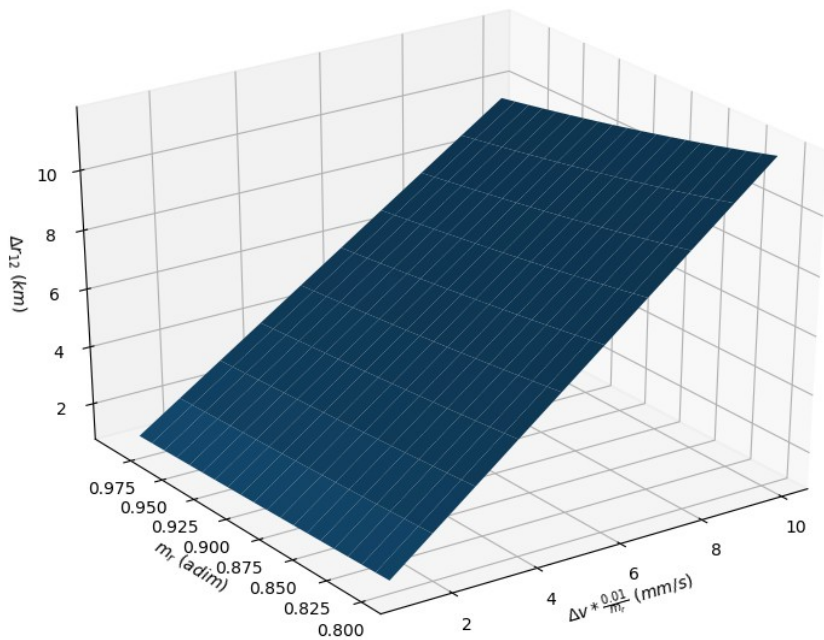

**Figure 1: Main asteroid final position variation for 2-Body Problem.**

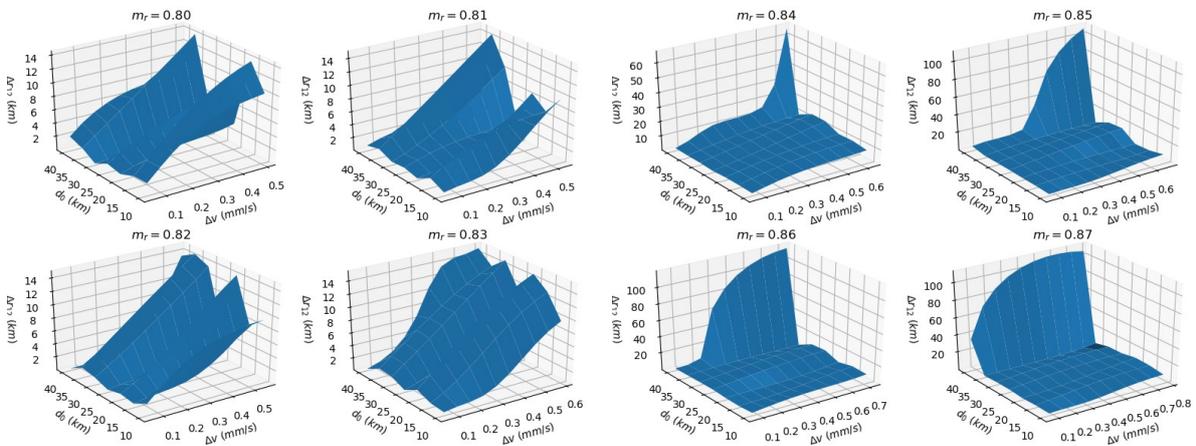

**Figure 2: Main asteroid final position variation (Mass ratio from 0.80 to 0.87).**

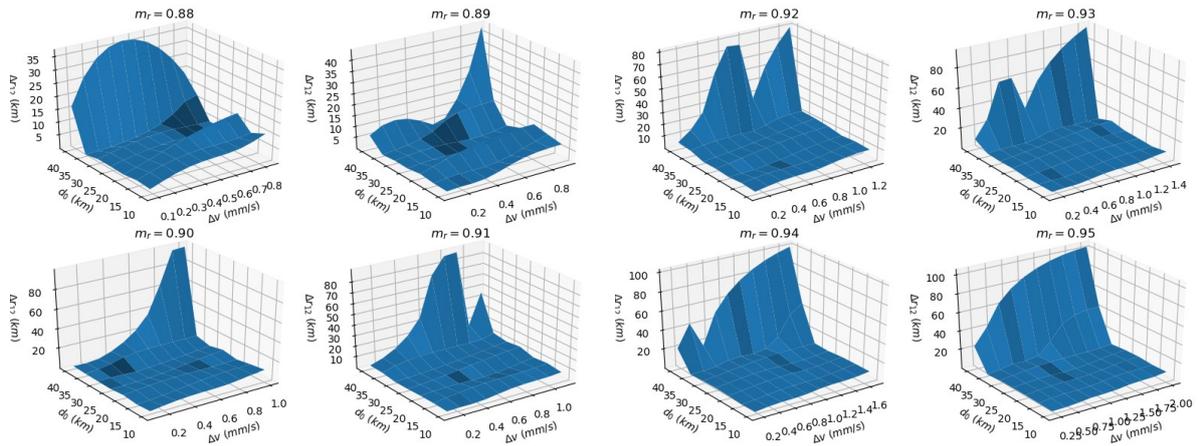

**Figure 3: Main asteroid final position variation (Mass ratio from 0.88 to 0.95).**

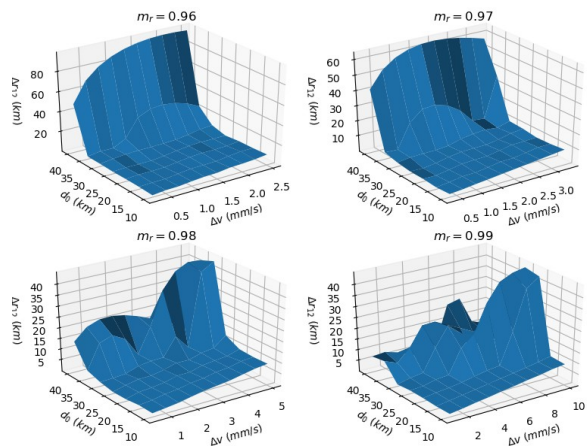

**Figure 4: Main asteroid final position variation (Mass ratio from 0.96 to 0.99).**

From Figures 2 to 4 it is possible to observe that larger $r_{12}$ values are obtained, in comparison with Figure 1. There is no visible relation between $r_{12}$ and $d_0$, but it is noticeable that the best values were obtained for the largest distances of $d_0 = 40$ km. In closer analysis of these simulations it was perceived that, in these situations, there was an escape of the secondary body from the main asteroid. The greatest deflections are a result of the effects of this gravitational interaction.

The best results are deflections of a little over a hundred kilometers. These are already approximately ten times better than the situations where the velocity increment is applied to the main body in the 2-Body Problem simulations, which show promising potential in the scenario where the deflection of a system of asteroids is necessary.

*3.2. Case 2*

Figures 5 to 9 presents the data in the same form as the previous figures, with an additional consideration that the four subplots on the left correspond to the lower bound initial distance $d_0 = 10$ km and the four subplots on the right correspond to the higher value of $d_0 = 40$ km. In turn, the Y-axis corresponds to the direction of the velocity increment.

Considering the variation along the Y-axis, of the direction of the velocity increment, it is possible to conclude that the best deflection scenarios occur when the increment is applied in the direction or the opposing direction of the movement ($\angle\Delta v$ equal to $0°$ or $180°$). The worst results happen for increments perpendicular to the movement ($\angle\Delta v$ equal to $90°$ or $270°$). This is a result that would be already expected in a 2-Body Problem dynamic scenario and to verify that it also happens in the 3-Body Problem dynamics only reinforces this pattern.

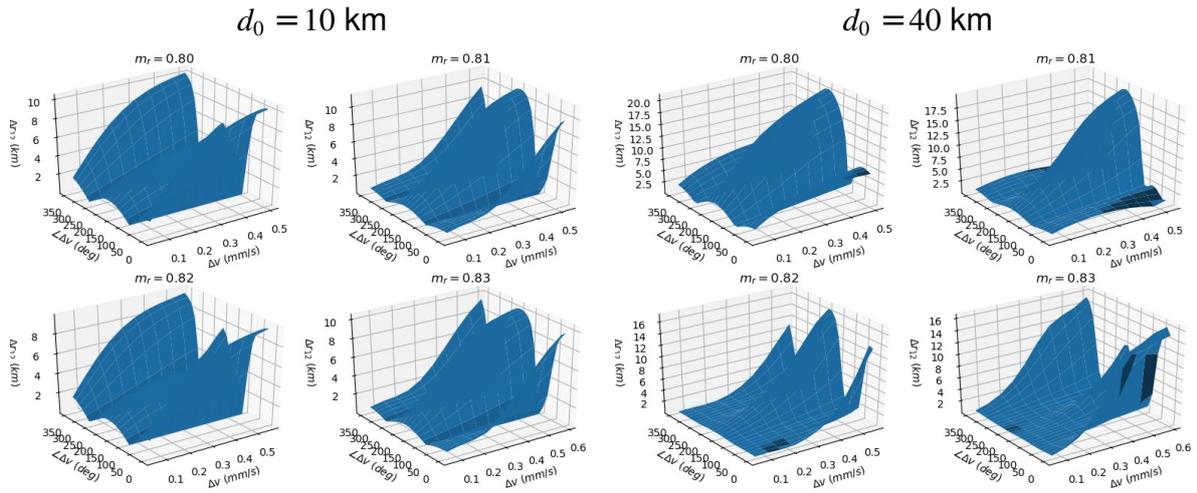

Figure 5: Main asteroid final position variation (Mass ratio from 0.80 to 0.83).

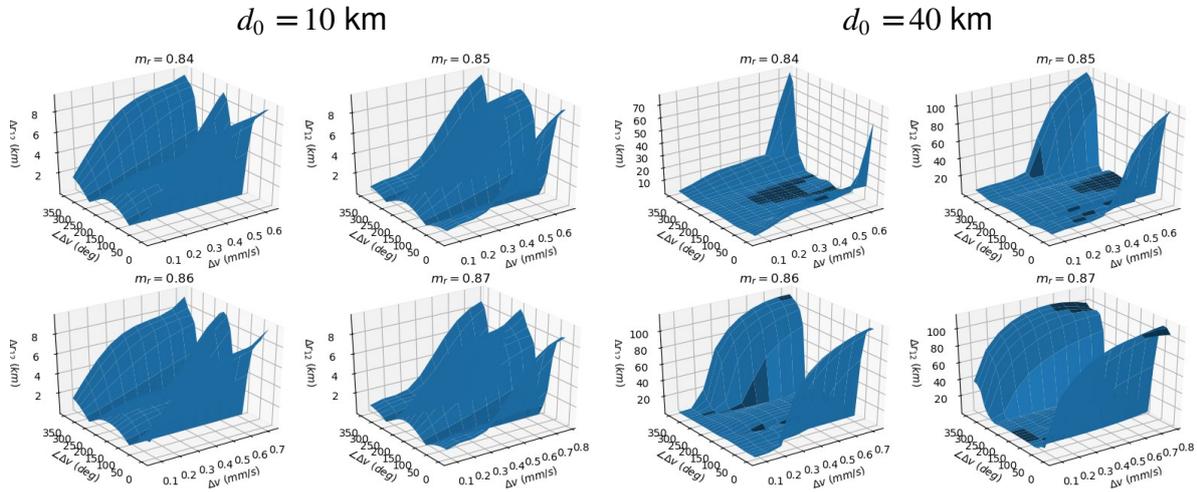

Figure 6: Main asteroid final position variation (Mass ratio from 0.84 to 0.87).

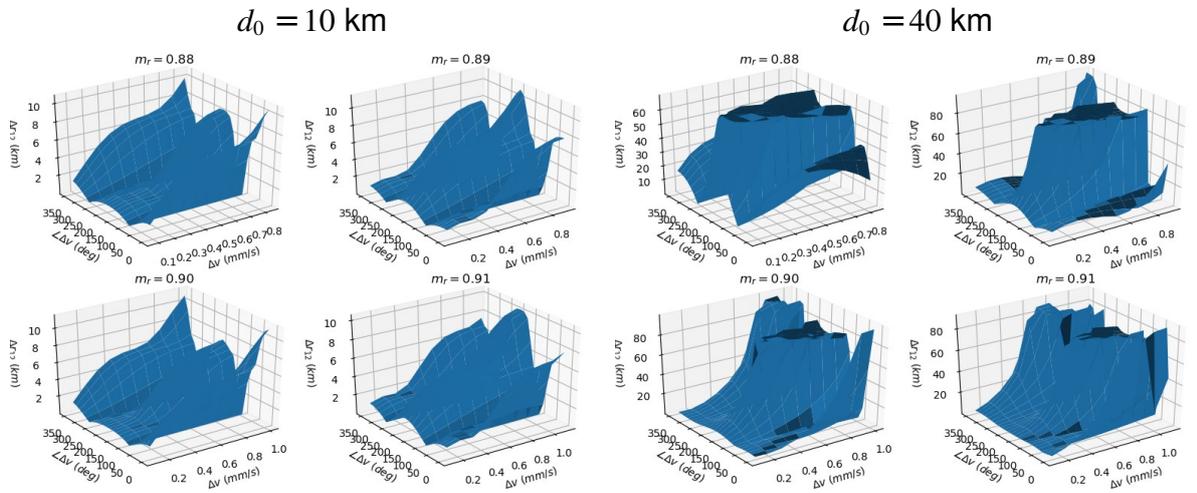

Figure 7: Main asteroid final position variation (Mass ratio from 0.88 to 0.91).

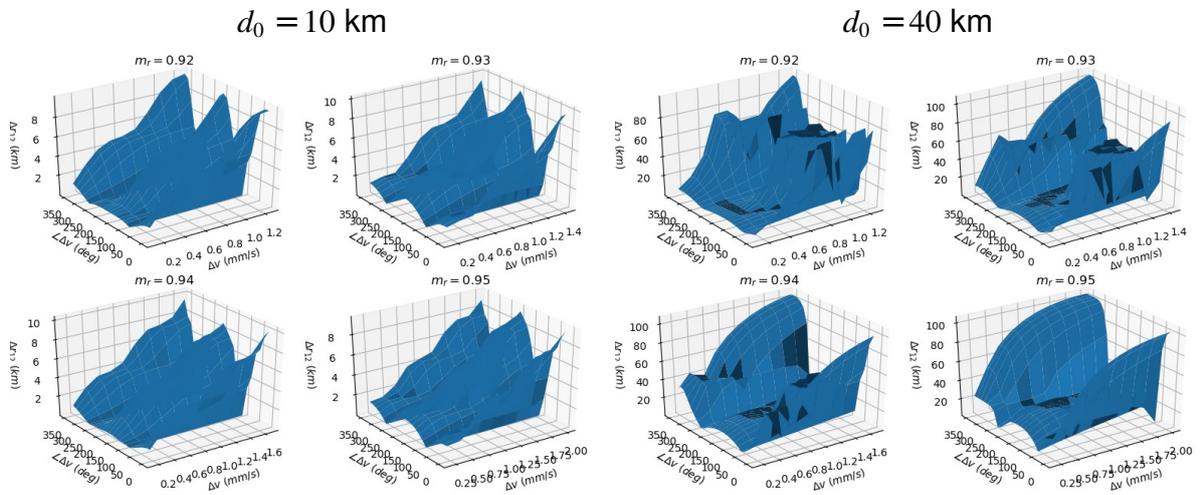

**Figure 8: Main asteroid final position variation (Mass ratio from 0.92 to 0.95).**

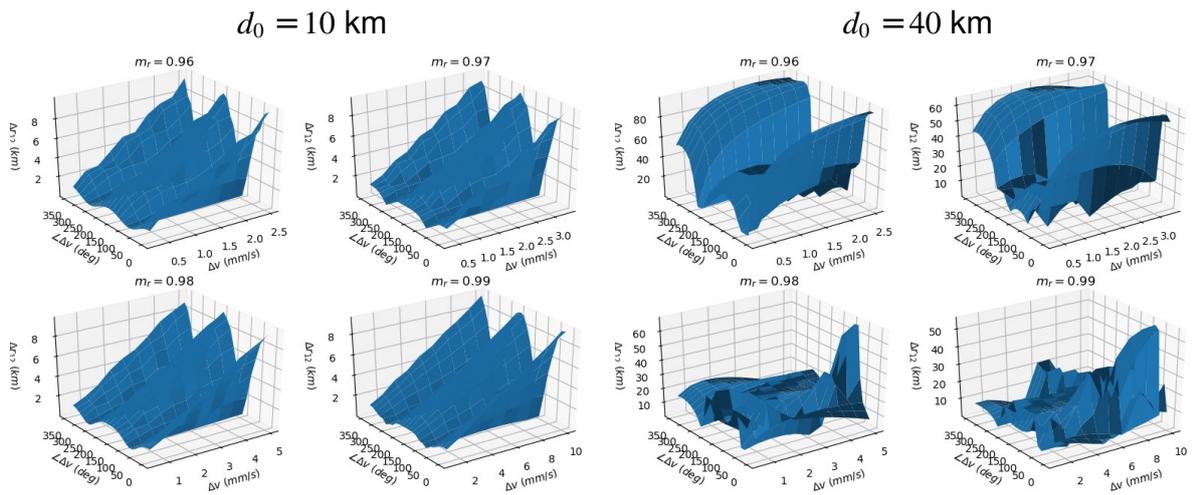

**Figure 9: Main asteroid final position variation (Mass ratio from 0.96 to 0.99).**

Once again, it is possible to verify results that deviate in one order of magnitude from other results in a same subplot. This can be explained with further analysis of each individual simulation, where an escape of the secondary body can be verified. This would also explain some of the plateaus seen, such as in Figure 7 for $d_0 = 40$ km and $m_r = 89°$ or $m_r = 90°$, where there is a limit in the deflection provided by the escape of the smaller body because, at one point, there is no longer a significant gravitational interaction between the bodies. These escapes scenarios were exclusively verified with the greatest initial distance $d_0 = 40$ km.

### 3.3. Case 3

Figures 10 to 14 present the data in the same manner as the previous figures. The only difference is the quantity presented along the Y-axis, which is the variation in the initial true anomaly of the system's orbit around the Sun ($f_0$).

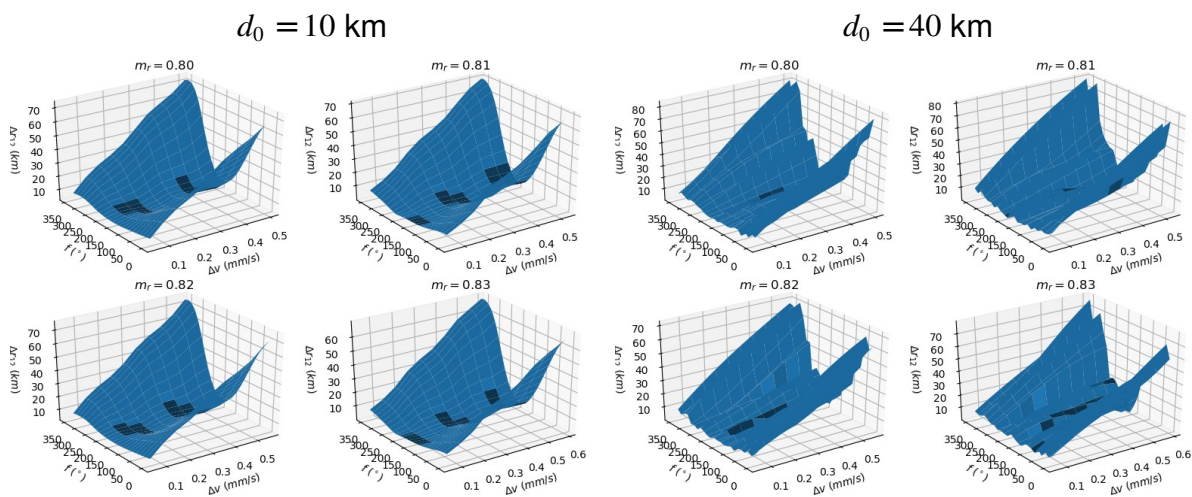

**Figure 10: Main asteroid final position variation (Mass ratio from 0.80 to 0.83).**

The results displayed the importance of the position of the system with a goal of deflecting the main asteroid, considering an increment in the velocity of the secondary body. Once again, this is a concept already expected for the orbit of a single asteroid in a 2-Body Problem, which is reinforced by being verified in the 3-Body Problem dynamic. Increments in the direction of movement, given near the perihelion, present greater deflections compared to increments given near the aphelion.

Once again, escape of the secondary body can be verified, this time in Figures 13 and 14, for $d_0 = 40$ km and $m_r > 94°$. A best case deflection of approximately $700$ km was obtained, indicating that a binary asteroid system with an heliocentric orbit with greater eccentricity is more susceptible to deflections compared to a circular orbit.

### 4. Conclusion

This study proposed an analysis of the effects of an increment in the velocity of the secondary body of a binary asteroid system. The variation of some of the character-istics of the system had the intention of understanding the influence of each one over the final results.

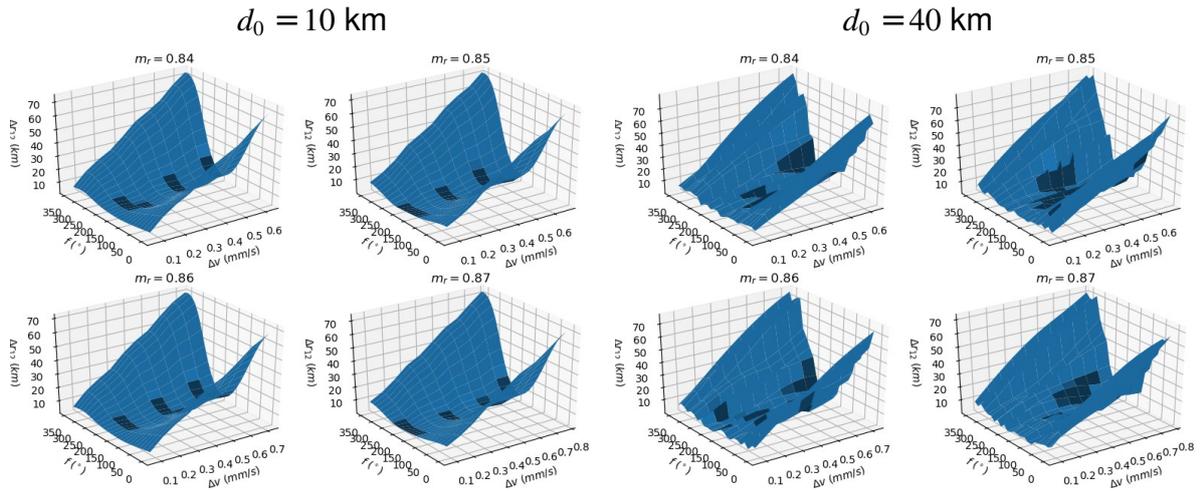

Figure 11: Main asteroid final position variation (Mass ratio from 0.84 to 0.87).

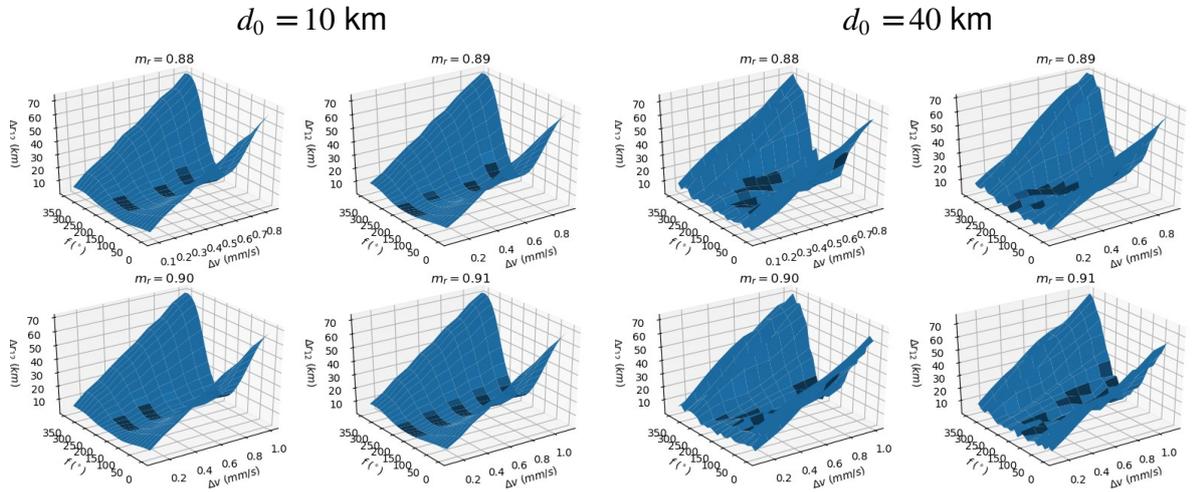

Figure 12: Main asteroid final position variation (Mass ratio from 0.88 to 0.91).

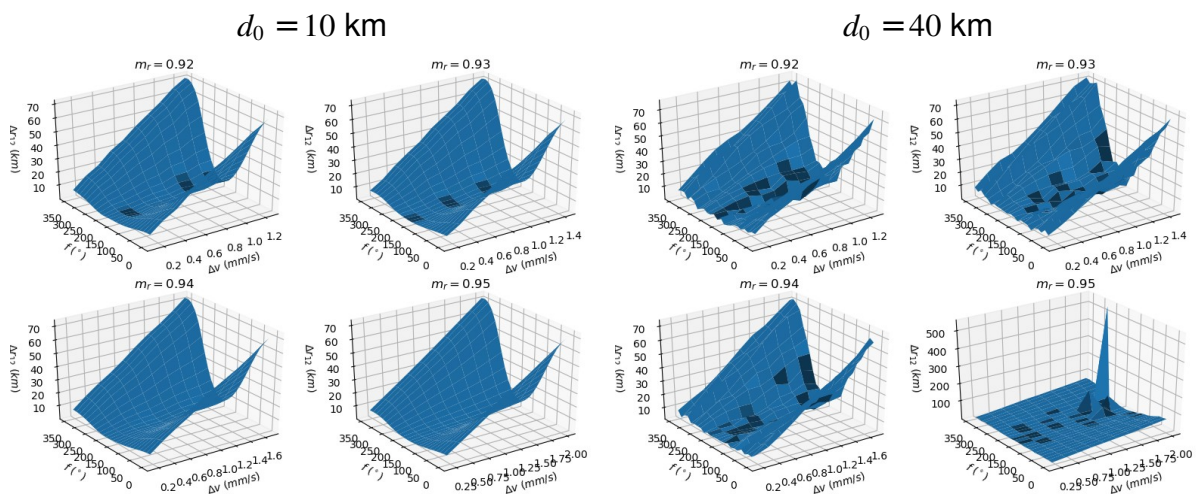

Figure 13: Main asteroid final position variation (Mass ratio from 0.92 to 0.95).

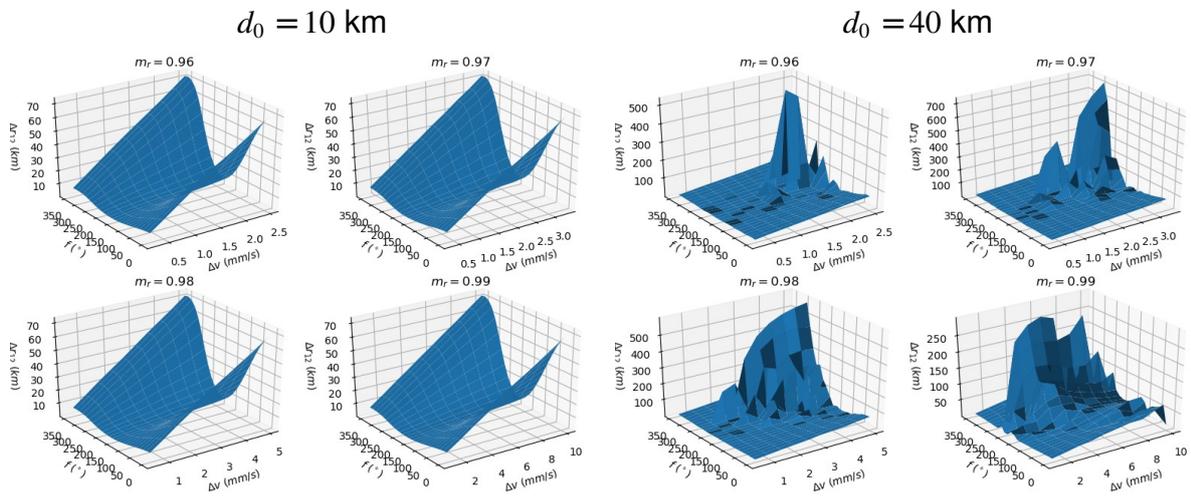

**Figure 14: Main asteroid final position variation (Mass ratio from 0.96 to 0.99).**

The deflection in the trajectory of the main asteroid ranges from a few kilometers to a couple hundred kilometers. Given the small increments in velocity, based on realistic projection of human missions, this result is to be expected. This values may not be enough to represent a successful deflection in a collision scenario. Nevertheless, they show a promising capacity to overcome future hazardous scenarios involving possible asteroid collisions. In addition, when comparing the 2-Body Problem dynamics with a single asteroid with the 3-Body Problem dynamics of an asteroid system, it was verified the advantage of incrementing the velocity of a smaller asteroid and taking advantage of the altered gravitational interaction between both bodies in achieving larger trajectory deflection.

It was verified that larger magnitudes of the velocity increment result in larger deflections, as expected. Results also benefit from a velocity increment directly in the direction or opposite to the direction of movement. Additionally, greater distances between the bodies of the asteroid system increase the chance of an escape of the secondary body, which in turn provides even greater deflections of the main body's trajectory. Finally, the mass ratio of the system did not show any direct influence in the magnitude of the deflection.

When applying the increments of velocity in a system with elliptical orbit, it was verified that the greatest deflections, at first, were achieved by incrementing the velocity near the perihelion region of the system's orbit. Nevertheless, increments near the aphelion region resulted in a larger number of escape of the secondary body situations, which was already verified as the situation that promotes the greatest deflections overall.

This study was successful in investigating some of the effects of a small increment in velocity of the secondary body. It still opens up the possibility of many more variations, from non-circular or inclined orbits to different total masses of the system and even conditions of an escape of the smaller body.

**Acknowledgments**

The authors would wish to express their appreciation for the support provided by the grants # 406841/2016-0 and 301338/2016-7 from the National Council for Scientific and Technological Development (CNPq), grants # 2018/19959-0 and 2016/24561-0 from São Paulo Research Foundation (FAPESP) and to the financial support from the National Council for the Improvement of Higher Education (CAPES / PrInt CAPES - INPE Project).